# Length Control and Frequency Tunability of Dipole Antennas Using High-Permittivity Material Termination for Ultrahigh-Field Magnetic Resonance Imaging

Aditya A. Bhosale[1], Yunkun Zhao[1], Divya Gawande[1], Komlan Payne[1], and Xiaoliang Zhang[1,2*]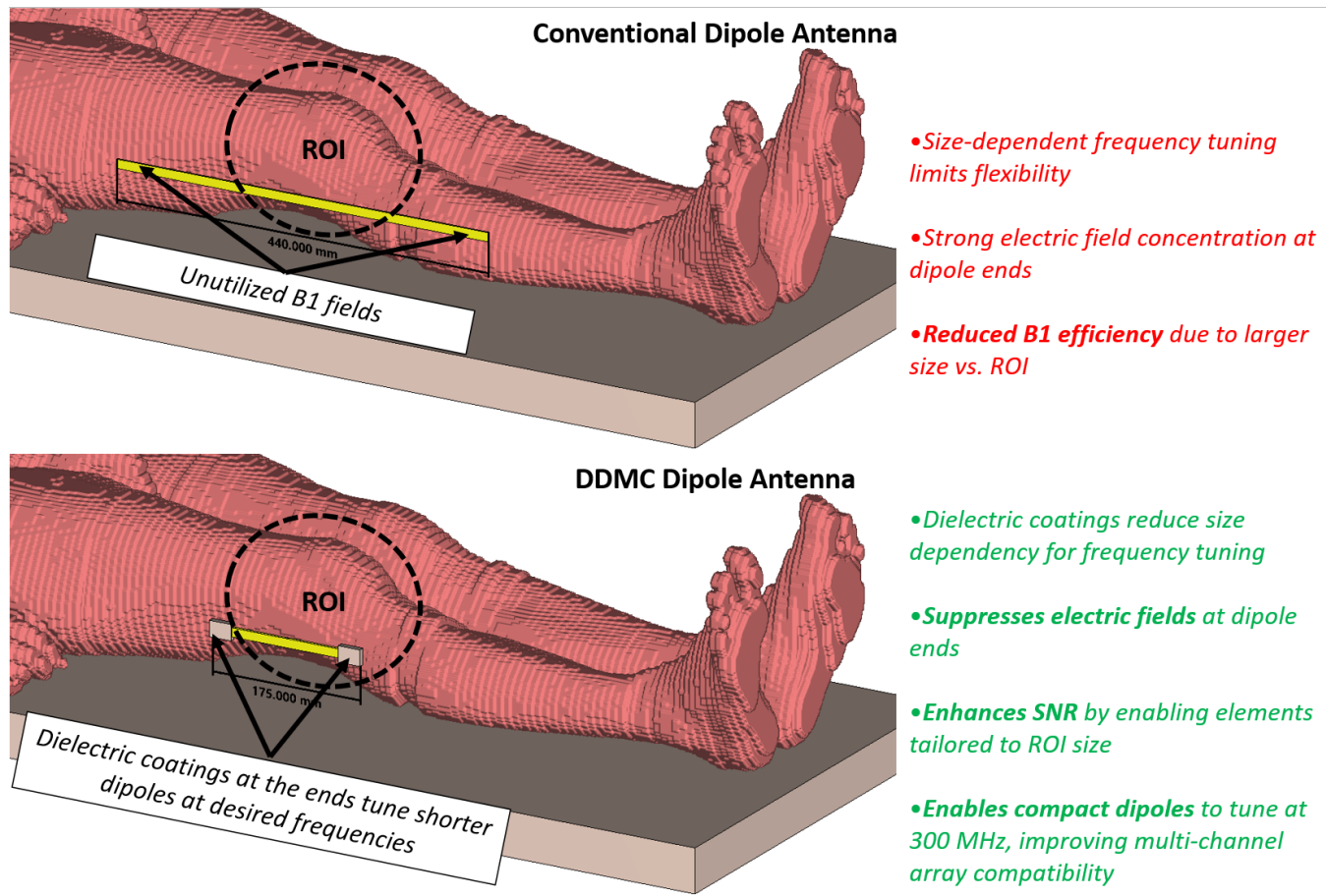

Caption: From Conventional to DDMC Dipole: Dielectric coatings modify a standard dipole, enabling compact tuning, reducing E-field concentration, and improving multi-channel compatibility at 300 MHz.

**Take-Home Messages**
- **Innovation:** The DDMC dipole antenna enables frequency tuning, reduces SAR, and improves B1 field uniformity without altering geometry.
- **Conclusion:** It enhances imaging safety, field uniformity, and SNR, making it ideal for high-field MRI.
- **Applications:** Designed for ultra-high-field MRI, particularly brain and knee imaging.
- **Breakthrough:** Overcomes dipole limitations with high-permittivity coatings for tunability, efficiency, and safety.
- **Highlights:** Demonstrates scalability for multi-channel MRI arrays with strong experimental validation.

# Length Control and Frequency Tunability of Dipole Antennas Using High-Permittivity Material Termination for Ultrahigh-Field Magnetic Resonance Imaging

Aditya A. Bhosale[1], Yunkun Zhao[1], Divya Gawande[1], Komlan Payne[1], and Xiaoliang Zhang[1,2*]

*Abstract* : This study presents a novel discretely dielectric material-coated (DDMC) dipole antenna design for ultra-high-field (UHF) MRI applications. This design improves frequency tuning, lowers electric field intensity, and reduces SAR by including discrete high-permittivity dielectric coatings at both ends of the dipole. The DDMC dipole's performance was compared to that of a fractionated dipole design using metrics such as inter-element coupling, B1 field distribution, and SNR. Simulations and experimental results showed that the DDMC dipole provided superior B1 field uniformity with significantly reduced B1 variation along the dipole conductor while reducing SAR, making it a safer and more efficient option for MR signal excitation and reception in UHF MR imaging. Furthermore, with its improved electromagnetic decoupling performance, the multichannel array made from the proposed DDMC dipoles shows promise for improving parallel imaging and imaging quality in UHF MRI, with future work focusing on material optimization and scalability for multi-channel arrays.

## I. INTRODUCTION

Conventional dipole antennas [1], [2] have gained popularity in ultra-high-field MRI applications [3]-[9] because of their simple fabrication, high-frequency capability, deeper penetration, and improved decoupling performance [10]-[16]. However, the frequency tuning of these antennas is determined solely by their physical length, with virtually no means to employ conventional tuning methods using lumped capacitors [17]-[19]. As a result, increasing the dipole length lowers the resonant frequency, making tunability strongly size-dependent. For 300 MHz applications, this limitation results in excessively long antennas, complicating their practical implementation and reducing their efficiency in many imaging applications, such as brain or knee imaging [20]-[22]. Additionally, the high electric fields generated at the ends of the dipole raise safety concerns, as they can lead to elevated specific absorption rate (SAR), further limiting their utility [23], [24].

Efforts to address these challenges have focused on shortening the dipole through techniques such as bending the arms or incorporating meander patterns to increase inductance [25]-[27]. These strategies allow for the tuning of shorter dipoles at higher frequencies by modifying the geometry, such as increasing the width or creating intricate patterns. Meanders, for example, can be applied at both ends, fractionated into segments, or distributed along the entire length of the conductor. Alternatively, the dipole arms can be folded outward to reduce the overall size. While these methods can achieve the desired frequency tuning, they still rely on inductance compensation through adjustments to the dipole's length. This dependence introduces additional complexities, as any frequency shift caused by loading requires further modifications to the antenna size. Furthermore, the altered geometry often disrupts the current distribution along the dipole, leading to irregularities in electromagnetic field distribution and signal performance [28]. The need for shorter dipoles is particularly pressing for applications involving small regions of interest, where longer dipoles result in significant field inefficiencies. Much of the field generated by larger antennas is wasted, leading to reduced filling factor and thus degraded imaging performance [29]. Longer dipoles also compromise penetration depth and increase the bulk of compact array systems, which limits their practicality in high-resolution imaging setups.

High dielectric materials might be a promising solution for overcoming these challenges. These materials are widely used in MRI applications for their ability to tune frequency through dielectric loading, manipulate the B1 field distribution, suppress electric fields, and lower SAR [12], [28], [30]-[37]. Building on these advantages, we propose a novel method that preserves the conventional geometry of dipole antennas while addressing their inherent limitations, particularly frequency tuning and SAR. Our design involves the addition of discrete dielectric coatings at both ends of the dipole. This approach enables frequency tuning without altering the antenna's physical length or geometry, ensuring the preservation of the current distribution. By strategically placing dielectric materials at

This work was supported in part by the NIH under a BRP grant U01 EB023829 and by the State University of New York (SUNY) under SUNY Empire Innovation Professorship Award.(Corresponding author: Xiaoliang Zhang.)

Aditya Ashok Bhosale, Yunkun Zhao, Divya Gawande, and Komlan Payne are with the Department of Biomedical Engineering, State University of New York at Buffalo, USA

Xiaoliang Zhang is with the Departments of Biomedical Engineering and Electrical Engineering, State University of New York at Buffalo, Buffalo, NY 14260 USA (e-mail: xzhang89@buffalo.edu).

the ends, where high electric fields are generated, the design not only achieves efficient frequency tuning but also reduces electric fields and lowers SAR, making it safer and more effective for 300 MHz applications. The performance of this novel design was evaluated against a popular fractionated dipole of the same length using both bench tests and 3D electromagnetic simulations. Bench tests assessed parameters such as B1 field distribution, electric field distribution, and coupling in a two-channel setup, while simulations analyzed B1 efficiency, signal-to-noise ratio, and SAR performance. These evaluations demonstrate the potential of the proposed design to provide an efficient, compact, and safe alternative to conventional dipole configurations in high-frequency MRI applications.

## II. METHODS

### A. Design Specifications

The dipole antenna designs for this study include both a fractionated dipole and a Discrete Dielectric Material Coated (DDMC) dipole. The fractionated dipole is 175 mm in total length, with a dipole strip width of 10 mm. The meander section of the dipole is 37.5 mm wide and 20 mm long. It is supported by a Teflon (PTFE) substrate with a dielectric constant of 2.1 and a thickness of 3.2 mm, with total substrate dimensions of 190 mm length and 50 mm width. Copper tape is applied to the top of the substrate to form the fractionated dipole conductors, and another layer of copper tape is added to the back of the substrate for shielding to reduce radiation loss. To match the impedance, a Cmatch capacitor is inserted in parallel with the coaxial feed port. The frequency tuning of the fractionated dipole is exclusively determined by the inductance introduced by the dipole conductors.

The DDMC dipole has a similar length (175 mm) and strip width (10 mm) and employs Teflon (PTFE) as the substrate material, with identical dielectric characteristics ($\varepsilon_r = 2.1$) and thickness (3.2 mm). The DDMC design distinguishes itself by including individual dielectric coatings at both ends of the dipole. The coating is 30 mm long, 22.5 mm wide, and 6 mm thick, composed of water-gelatin gel, and has a dielectric constant of 78. Copper tape is applied to the top of the substrate to form the dipole conductors, while additional copper tape is added to the back to provide copper shielding and limit radiation loss. To achieve impedance matching, a Cmatch capacitor is connected in parallel to the coaxial feed port. The DDMC dipole is tuned using both the dipole structure's inductance and the dielectric coatings' attributes (dielectric constant, thickness, length, and width). Unlike the fractionated dipole design, this tuning method enables for fine-tuning of shorter dipoles while keeping the conductor dimensions unchanged. Figure 1 depicts the simulation models of the fractionated and DDMC dipoles, with the dimensions annotated for clarity.

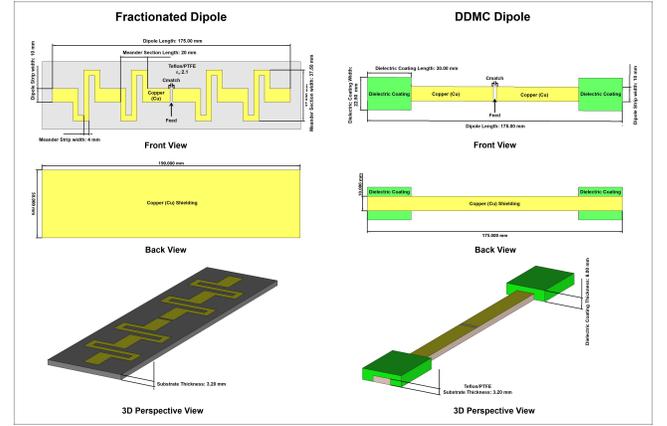

Fig. 1. Simulation models of the fractionated dipole and the proposed DDMC dipole. The left section illustrates the fractionated dipole, including the top-left front view with labeled dimensions, the middle-left back view showing copper shielding, and the bottom-left 3D perspective of the overall design. The right section depicts the DDMC dipole, highlighting the top-right front view with dielectric coatings and labeled specifications, the middle-right back view with copper shielding and component layout, and the bottom-right 3D perspective of the complete design.

### B. Fabrication Process

The fabrication of the DDMC dipole involves several key steps. A Teflon (PTFE) substrate with the specified dimensions (175 mm × 10 mm, thickness 3.2 mm) was first prepared as the base structure for the dipole. Copper tape was applied to form the dipole conductors, followed by an additional layer of copper tape on the back of the substrate for shielding to minimize radiation loss. The coaxial feed was integrated, and a Cmatch capacitor was added in parallel to the feed port for impedance matching.

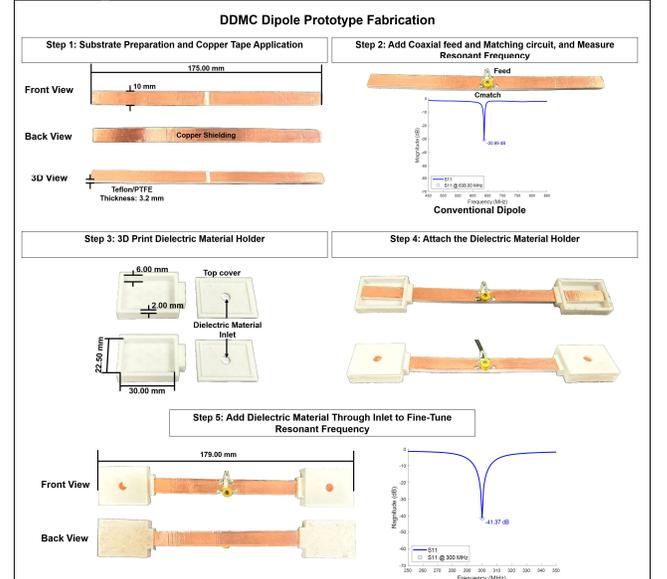

Fig. 2. The fabrication process of the DDMC dipole prototype. Steps include substrate preparation, copper tape application, coaxial feed and matching circuit integration, 3D printing and attachment of the dielectric material holder, and final tuning by adding dielectric material.

A 3D-printed holder was designed and attached to precisely position the dielectric material at both ends of the dipole. The dielectric material, a water-gelatin gel ($\varepsilon_r = 78$), was applied evenly to the ends of the dipole. The dielectric coatings measured 30 mm in length, 22.5 mm in width, and 6 mm in thickness at both ends. The antenna was tuned by adjusting the amount of the dielectric material to achieve the desired resonant frequency. Figure 2 illustrates the fabrication process, including substrate preparation, copper tape application, coaxial feed integration, 3D printing, and dielectric material positioning.

### C. Simulation Setup

Both frequency-domain (finite element) and time-domain voxel simulations were conducted using CST Microwave Studio to evaluate the B1 field, electric field distributions, B1 efficiency, signal-to-noise ratio (SNR), and specific absorption rate (SAR). Frequency-domain simulations were used to evaluate the B1 field and electric field distributions for both dipole designs under unloaded conditions. Open boundary conditions were applied with outer walls placed λ/4 away from the model. These simulations were performed with very fine mesh sizing for accurate results. For the time-domain voxel simulations, the Hugo human knee voxel model was used, with a resolution of 1x1x1 mm³. The voxel model included various tissue types such as blood, bone cortical, bone spongiosa, fat, muscle, and skin, modeled with material properties for 300 MHz. The first channel of each dipole type was positioned 80 mm away from the center of the human knee model, while the second channel was placed at a 45° rotation relative to the position of the first channel.

The voxel simulations evaluated B1 efficiency in micro-Tesla/sqrt(accepted power) and SNR by dividing B1 efficiency by the calculated noise value. The noise calculation was based on losses from electric fields and losses in all materials included in the simulations, including lossy materials such as dielectric materials and tissue materials in the voxel, as well as all metals used in the simulation. SAR was calculated following the IEEE/IEC 62704-1 standard, using a 10-g averaged volume with mass accuracy of 0.0001% normalized to accepted power. The coaxial feed was represented using a lumped port in the simulations, and impedance matching was achieved by modeling a lumped element capacitor parallel to the feed point. The frequency-domain simulations involved adaptive meshing with a maximum of 8 passes to ensure convergence. All simulations were repeated five times for each simulation task and case to verify the repeatability and the authenticity of the results. The simulation times for each run were approximately 2 hours for frequency-domain simulations and 5 hours for voxel time-domain simulations.

### D. Bench Test Setup

The bench test measurements were performed to validate the simulation results under unloaded conditions. The following equipment was used: Network Analyzer: Keysight E5061B; Probes: In-house fabricated 2 cm loop H-field probe and E-field probe; Positioning System: High-precision 3D positioning system constructed using a CNC router (Genmitsu PROVerXL 4030). The probes were calibrated using an in-house LC loop calibration system to ensure field normalization based on power outputs calculated from S-parameters. The prototype setups of the fractionated dipole and DDMC dipole used for these measurements are shown in Figure 3. The B1 field measurements were conducted in the XZ and YZ planes. For the B1 field measurements, the B1x and B1y fields were measured by positioning the probes perpendicular to the respective axes. The B1xy fields were obtained by combining B1x and B1y fields using the root mean square method. The B1 field measurement setups and corresponding field distributions in the XZ and YZ planes for both dipole designs are illustrated in Figures 4 and 5, respectively.

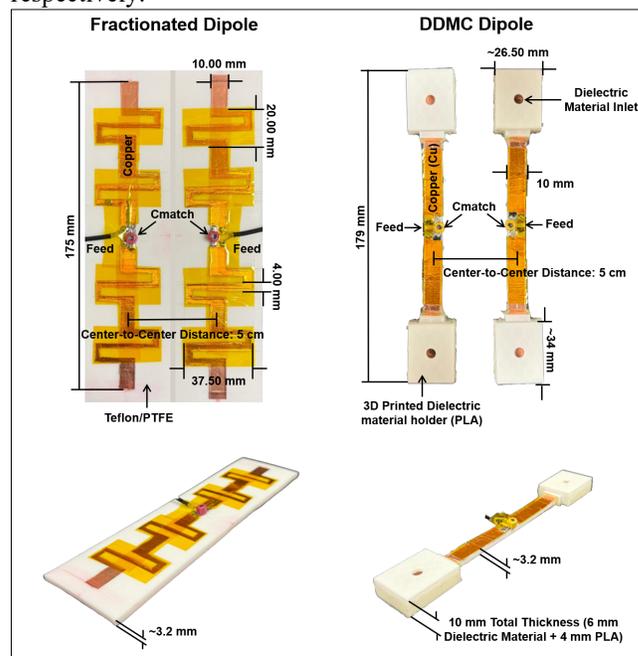

Fig. 3. Prototype setups of the fractionated dipole and DDMC dipole for coupling evaluations. The top-left panel shows the two-channel fractionated dipole setup with dimensions and specifications at a center-to-center distance of 5 cm. The top-right panel displays the corresponding DDMC dipole two-channel setup. Both setups were used to evaluate coupling as the center-to-center distance was varied from 5 cm to 10 cm. The bottom-left and bottom-right panels present the 3D perspective views of the fractionated dipole and DDMC dipole, respectively.

The E-field measurements were also performed in the XZ plane, and the measurements were conducted 20 mm and 30 mm above the coils for the XZ and YZ planes, respectively. The E-field measurement setup and the measured electric field distributions in the XZ plane for both dipole designs are presented in Figure 6. The S-parameters (S11, S21) were used to assess inter-element isolation and coupling between the two dipole channels. To evaluate the coupling performance, the two dipole channels were placed at center-to-center distances ranging from 5 cm to 10 cm. The coupling evaluation setups for the fractionated dipole and DDMC dipole are detailed in Figure 3. The inter-element isolation was assessed using S-parameters, specifically S11 and S21. The isolation was considered satisfactory when S21 values were lower than -20 dB, indicating well-decoupled behavior. No decoupling circuits were used, and both dipole types exhibited good decoupling behavior under the evaluated conditions.

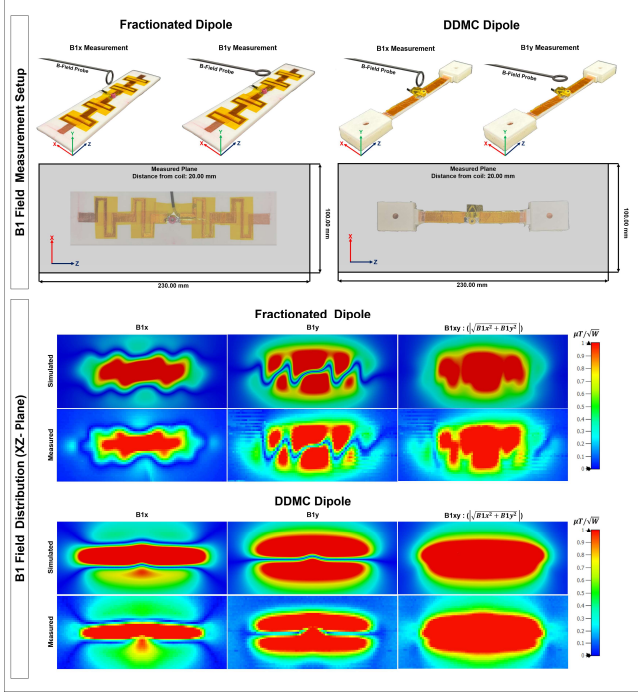

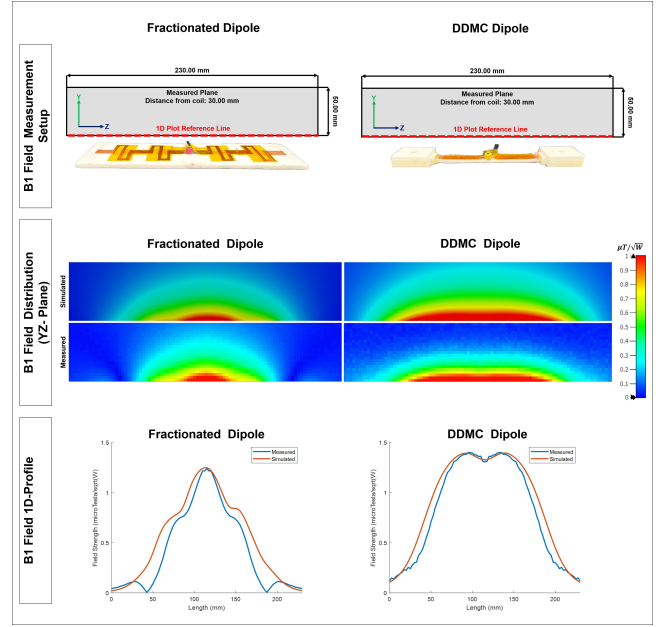

Fig. 4. B1 field distribution in the XZ plane for the fractionated and DDMC dipoles. The top section illustrates the measurement setup used for the XZ plane B1 field distribution. The left part shows the measurement setup for the fractionated dipole, and the right part shows the setup for the DDMC dipole. The measurements included B1x (probe perpendicular to the X-axis) and B1y (probe perpendicular to the Y-axis), both taken at a distance of 20 mm from the coil, covering an area of 230 mm x 100 mm for both dipoles. The bottom section presents the B1 field distribution in the XZ plane for both dipoles. The top part of the bottom section shows the field distribution for the fractionated dipole, while the bottom part shows the distribution for the DDMC dipole. Simulated field distributions are compared with the measured data from the setup shown in the top section. B1x and B1y fields were combined using the root mean square method to form the B1xy field distributions.

*E. Dielectric Material*

The dielectric material used for the coatings at the ends of the DDMC dipole was a solution of distilled water and gelatin, which was poured into the dielectric material holders via the dedicated inlet. The material was applied evenly to both ends of the dipole until the desired frequency tuning was achieved. To address discrepancies between the dielectric constant in bench tests and simulations (due to material variations), the dielectric material holders were fabricated with extra thickness to compensate for these differences.

*F. Data Handling and Results Processing*

B1 efficiency, SNR, and SAR field distributions were directly exported from CST Microwave Studio. The SNR heatmaps were generated by recording the average SNR in each region of interest (ROI) block for each plane, with the matrix then exported to MATLAB for plotting.

Fig. 5. B1 field distribution in the YZ plane for the fractionated and DDMC dipoles. The figure is divided into three sections: The top section shows the B1 field measurement setup, with a 3D side view of each dipole and the measured plane positioned 30 mm above each dipole. The plane measures 230 mm x 50 mm, and a 1D reference line is marked for plotting the B1 field 1D profile. The middle section presents the B1 field distribution in the YZ plane. Simulated and measured field distributions are compared for both dipole designs, with the left side showing the fractionated dipole and the right side showing the DDMC dipole. The bottom section displays the B1 field 1D profile, where the measured line plot is compared with the simulated line plot. The left part shows the 1D profile for the fractionated dipole, and the right part shows the profile for the DDMC dipole.

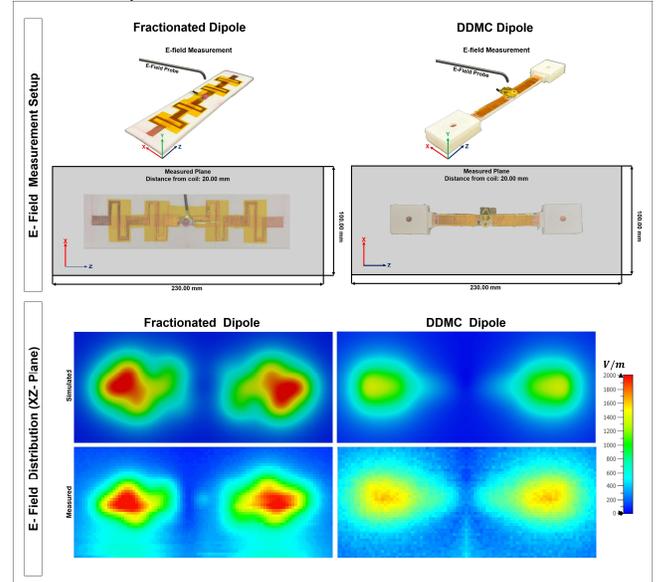

Fig. 6. Measured electric field distributions in the XZ plane compared with simulated distributions for the fractionated and DDMC dipole designs. The top section shows the electric field measurement setup, with the left side depicting the setup for the fractionated dipole and the right side for the DDMC dipole, including the orientation of the E-field probe and the measured plane (230 mm x 100 mm) positioned 20 mm away from the coil. The bottom section compares the measured and simulated E-field distributions, with the left side showing data for the fractionated dipole and the right side for the DDMC dipole.

## III. RESULTS

### A. Inter-element coupling and Q-factor evaluations

The measured inter-element coupling evaluations for the fractionated dipole and DDMC dipoles are shown in Figure 7. The measurements were conducted using the setup illustrated in Figure 3, with center-to-center distances ranging from 5 cm to 10 cm. The figure presents the S-parameters for each distance case, where S11 (reflection coefficient) is used to display the tuned and matched frequency, and S21 (transmission coefficient) indicates crosstalk or interference between the two channels in the setup. For the fractionated dipole, the S21 values ranged from -31.45 dB to -51.78 dB, with decoupling improving as the center-to-center distance increased. In contrast, the DDMC dipole showed slightly higher coupling across all distances, with S21 values ranging from -31.99 dB to -43.56 dB, suggesting that while the DDMC dipole provides adequate isolation, it exhibits a higher degree of coupling compared to the fractionated dipole.

The Q-factor, calculated based on the center frequency and the -3dB bandwidth of the S21 plots, was significantly higher for the fractionated dipole, approximately 63.15, compared to the DDMC dipole, which had a Q-factor of around 16.66. This higher Q-factor for the fractionated dipole indicates lower losses and a narrower resonance bandwidth.

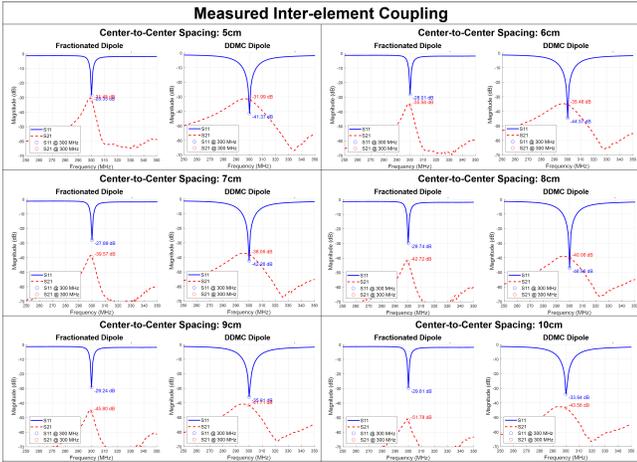

Fig. 7. Measured inter-element coupling evaluations for the fractionated and DDMC dipoles. The measurements were conducted using the setup shown in Figure 3, with center-to-center distances ranging from 5 cm to 10 cm. The figure presents the S-parameters for each distance case, showing S11 (reflection coefficient) to display the tuned and matched frequency, and S21 (transmission coefficient) to indicate crosstalk/interference between the two channels in the setup.

### B. B1 Field Distribution: Measured vs. Simulated

The B1 field distributions for the fractionated dipole and DDMC dipole were evaluated in the XZ and YZ planes, with results compared between measurements and simulations. Figures 4 and 5 illustrate the B1 field distributions in the XZ and YZ planes, respectively. The B1xy field, calculated as the root mean square of the B1x and B1y components, provides an overall representation of the field distribution for both dipole designs. In the XZ plane (Figure 4), the fractionated dipole exhibited B1 fields concentrated primarily at the center, with significantly lower field strengths near the ends. This uneven distribution was attributed to the meanders in the fractionated dipole design, which caused ineffective or lost current near the ends despite its overall larger size compared to the DDMC dipole. The B1 field coverage spanned approximately 71% of the fractionated dipole's total length. Conversely, the DDMC dipole displayed a much more uniform B1 field distribution along its length, resembling the conventional dipole's current distribution. The B1 field coverage spanned approximately 96% of the DDMC dipole's length, indicating better field uniformity compared to the fractionated dipole. The uniform current distribution of the DDMC dipole contributed to its symmetric and even B1 field distribution, which contrasts with the non-uniform and asymmetric field distribution of the fractionated dipole caused by its meanders.

In the YZ plane (Figure 5), similar trends were observed. The fractionated dipole showed B1 field concentrations near the center, spanning only 62% of the dipole's length. The 1D profile further confirmed this, revealing signal drops near the meanders, consistent with the non-uniform current distribution observed in the XZ plane. The DDMC dipole, on the other hand, demonstrated uniform B1 field coverage in the YZ plane, spanning approximately 86% of its total length. The 1D profile confirmed this improved coverage and field distribution, highlighting the DDMC dipole's ability to achieve enhanced field uniformity and penetration. Moreover, the DDMC dipole generated a half-wavelength behavior typical of conventional dipole designs, with improved field penetration and uniformity compared to the fractionated dipole. Additionally, the simulated results and measured results were in strong agreement, confirming the accuracy of the simulations in predicting the B1 field distributions and the trends observed during experimental evaluation.

### C. Electric Field Distribution: Measured vs. Simulated

The electric field distributions of the fractionated dipole and DDMC dipole were evaluated using both simulations and measurements, as shown in Figure 6, which presents the results on the XZ plane. The peak simulated electric field strength for the fractionated dipole was 2178 V/m, while the measured value was 2247 V/m. For the DDMC dipole, the peak simulated electric field strength was 1426 V/m, and the measured value was 1623 V/m. The DDMC dipole demonstrated a 32% reduction in average peak electric field strength compared to the fractionated dipole.

The electric fields were stronger near the ends for both designs, but the DDMC dipole consistently generated lower peak and overall electric field distributions compared to the fractionated dipole. This reduction in electric field strength would improve imaging safety due to lower SAR levels. Additionally, as electric fields contribute to losses, the lower electric fields generated by the DDMC dipole are expected to positively impact SNR. The measured electric field distributions and values (trends) were in great agreement with the simulations.

### D. B1 Efficiency: Voxel Simulations

The B1 efficiency distribution for both the fractionated dipole and DDMC dipole was evaluated through voxel simulations in the sagittal (YZ), coronal (XZ), and axial (XY) planes at various Z positions, as shown in Figure 8. In the sagittal YZ plane, the fractionated dipole showed higher field strength at the center due to its concentrated current distribution, but this dropped near the ends due to the meander-induced non-uniform current distribution. In contrast, the DDMC dipole exhibited a more uniform B1 efficiency along its length, resulting in higher overall efficiency. The coronal XZ plane further confirmed these observations, with the DDMC dipole showing a more consistent field distribution, while the fractionated dipole had significant drop-offs near its ends. In the axial XY plane, the DDMC dipole demonstrated significantly better coverage, resulting in improved penetration and signal strength, crucial for higher SNR and enhanced imaging quality in MRI applications. Overall, the B1 efficiency profiles across all planes indicated that the DDMC dipole provides a more uniform and consistent field distribution, which is advantageous for achieving improved SNR and image quality, particularly in high-resolution imaging scenarios.

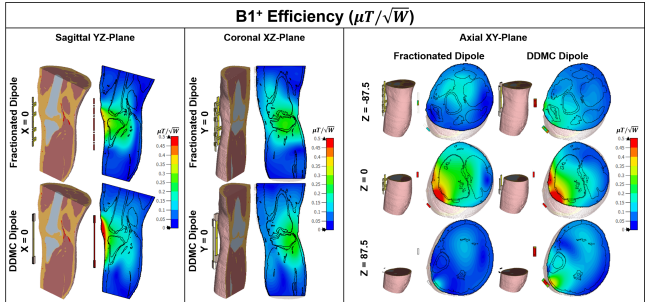

Fig. 8. B1+ efficiency distributions in the sagittal (YZ-plane), coronal (XZ-plane), and axial (XY-plane) planes, evaluated in voxel simulations using the human knee voxel model and a two-channel setup for both dipole types. The left section shows the sagittal YZ-plane field distributions, with the fractionated dipole (top) and DDMC dipole (bottom) results. The middle section presents the coronal XZ-plane field distributions for the fractionated dipole (top) and DDMC dipole (bottom). The right section displays the axial XY-plane field distributions for the fractionated dipole (left) and DDMC dipole (right), evaluated at three different precision levels for each dipole type.

### E. Signal-to-Noise Ratio: Voxel Simulations

The Signal-to-Noise Ratio (SNR) for the fractionated dipole and DDMC dipole was assessed through voxel simulations using a human knee voxel bio-model, as shown in Figure 9. SNR evaluation focused on region-of-interest (ROI) matrices in the sagittal, coronal, and axial planes to analyze field distribution attributes such as central coverage, performance near the dipole ends, penetration, and overall uniformity. In the sagittal plane, which included a 3×5 grid of ROIs, the fractionated dipole exhibited strong SNR near the center of the dipole (0.69) but significant drop-offs near the ends (0.09). This non-uniformity was caused by uneven current distribution along the dipole length, leading to limited coverage and diminished imaging performance at the extremities. Conversely, the DDMC dipole achieved a more balanced distribution, with consistently high SNR values near the center (0.96) and improved SNR near the ends (0.30). This uniformity enhanced coverage and ensured reliable performance along the dipole's entire length. The average sagittal SNR was 0.294 for the fractionated dipole and 0.414 for the DDMC dipole, reflecting an improvement of 40.82%.

In the coronal plane, also consisting of a 3×5 grid of ROIs, the fractionated dipole demonstrated its highest SNR values at the center (0.41) but experienced a decline toward the periphery, with values as low as 0.09. This uneven distribution reduced coverage and field strength in peripheral regions. In contrast, the DDMC dipole provided a more uniform SNR profile, with central values of 0.45 and consistently higher peripheral values (0.36 and 0.32), contributing to improved field coverage and better overall performance. The average coronal SNR was 0.231 for the fractionated dipole and 0.327 for the DDMC dipole, resulting in a 41.21% improvement. The axial plane featured a cross-pattern ROI with five points arranged in a plus shape. Here, the fractionated dipole exhibited strong central SNR (0.70) but weaker and inconsistent peripheral values (0.42, 0.33), indicating limited penetration and uneven coverage. The DDMC dipole displayed higher central SNR (1.00) and more consistent peripheral values (e.g., 0.47, 0.64), offering broader and more uniform field distribution with better penetration. The average axial SNR was 0.414 for the fractionated dipole and 0.522 for the DDMC dipole, marking an improvement of 26.09%. The combined analysis across all planes indicated an overall average SNR of 0.313 for the fractionated dipole and 0.421 for the DDMC dipole, representing an overall improvement of 34.42%. The DDMC dipole consistently outperformed the fractionated dipole by achieving superior central coverage, improved SNR near the dipole ends, enhanced field penetration, and greater uniformity along its length and across its width.

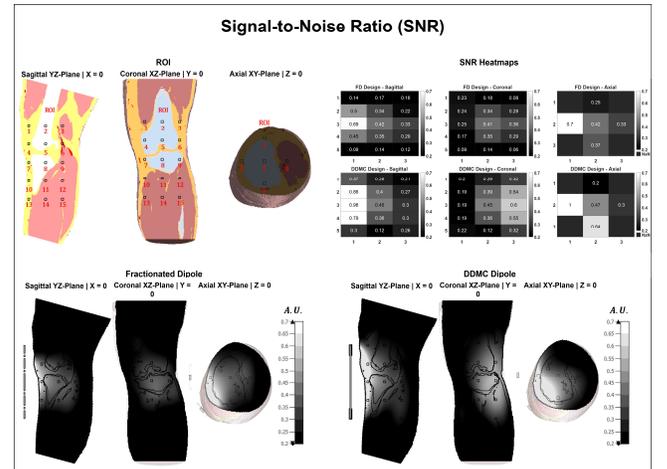

Fig. 9. SNR evaluation in voxel simulations using the human knee voxel model. The top left panel illustrates the human knee voxel bio-model with defined SNR regions of interest (ROIs) in each plane. The sagittal and coronal planes include a 3x5 grid of 5 mm x 5 mm ROIs, while the axial

plane features a cross-pattern ROI consisting of five points arranged in a plus-shaped pattern. The top right panel presents heatmaps comparing the SNR in the ROIs for the fractionated and DDMC dipoles, where each block represents the average SNR in the corresponding ROI of the respective plane. The bottom left panel displays the SNR distribution in the central sagittal, coronal, and axial planes for the fractionated dipole, while the bottom right panel shows the corresponding SNR distribution for the DDMC dipole.

*F. Specific Absorption Rate: Voxel Simulations*

Figure 10 presents the 10-gram averaged Specific Absorption Rate (SAR) plots for the fractionated dipole (left) and DDMC dipole (right). The peak SAR value for the fractionated dipole is 3.9607 W/kg, while the peak SAR value for the DDMC dipole is 0.8904 W/kg. This results in a significant percentage difference of approximately 77.52%, indicating that the DDMC dipole has a substantially lower SAR distribution compared to the fractionated dipole.

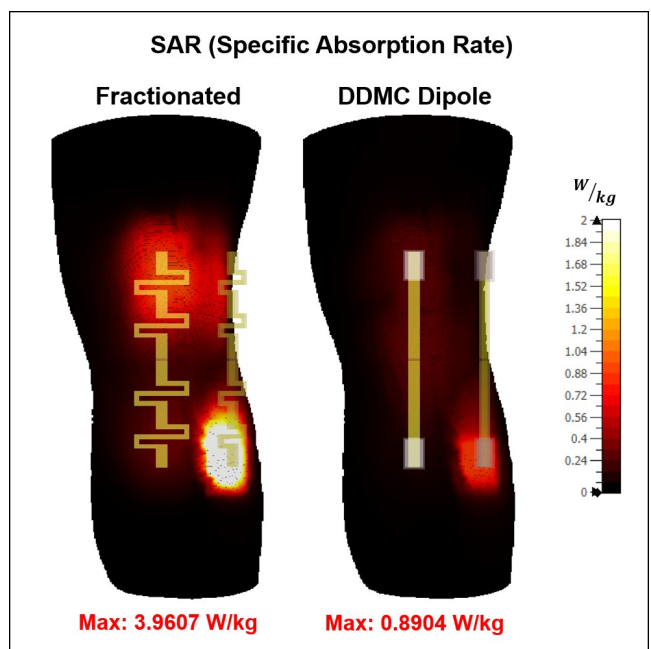

Fig. 10. 10-gram averaged specific absorption rate (SAR) plots for the fractionated (left) and DDMC dipole (right).

IV. DISCUSSIONS & CONCLUSION

In this study, a discretely dielectric material-coated dipole antenna is proposed and successfully developed for MR signal excitation and reception at ultrahigh magnetic fields. The design offers an efficient solution for frequency tuning without compromising the physical length or field distribution of the dipole antenna, while also enhancing imaging safety by reducing electric fields. This approach allows the physical length of the dipole antenna to be tailored to meet the requirements of specific imaging applications with improved filling factor and enhanced detection efficiency.

The design is validated through a comparison between two dipole designs, fractionated and DDMC, by analyzing their effects on inter-element coupling, B1 field distribution, electric field distribution, B1 efficiency, SNR, and SAR. Both designs were well decoupled over the distances tested, with coupling values of -20 dB or higher, indicating effective decoupling. The fractionated dipole performed slightly better over longer distances. However, given the size of the fractionated dipole, the distances tested were not particularly close. For the knee imaging setup, both dipole designs performed excellently at the tested distance, which was based on an 8-channel spacing for a two-channel setup.

The DDMC dipole had a more uniform B1 field distribution and better B1 coverage (71% to 96% in the XZ plane and 62% to 86% in the YZ plane), resulting in higher SNR and image quality. The inclusion of high-dielectric coatings in the DDMC dipole resulted in a 32% reduction in electric field strength when compared to the fractionated dipole. While the addition of dielectric materials reduced the Q-factor, potentially lowering SNR, the improved field uniformity and reduced electric field intensity compensated, resulting in better overall SNR and reduced SAR. Minimal changes to the conductor design helped maintain uniformity in both the B1 and current distribution, resulting in efficient B1 field generation. SAR measurements confirmed that the DDMC dipole lowers peak SAR, making it safer for high-field MRI applications. While the fractionated dipole provided superior decoupling and a higher Q-factor, it also resulted in a higher SAR due to its stronger electric field distribution. As a result, the DDMC dipole outperforms in terms of field uniformity, SNR, SAR, and safety, making it a better choice for high-resolution MRI systems.

Based on simulation results, a water-gel solution was chosen for the prototype; however, more practical dielectric materials could be used to achieve comparable performance. Using materials with a higher relative permittivity could further reduce the design's size, increasing its compactness. Future research will focus on optimizing coating dimensions and investigating higher-permittivity materials, which would be useful for ultra-high-field MRI systems. Further research could investigate optimizing dielectric materials to find cost-effective options for broader applicability. The scalability of the DDMC dipole could also be investigated for integration into multi-channel array systems, enhancing parallel imaging capabilities and making the design more adaptable for clinical and research use.